\documentclass[11pt]{article}
\usepackage{graphicx}
\usepackage{amssymb}
\usepackage{amscd}
\usepackage{mathrsfs}
\usepackage{longtable,lscape}
\usepackage{amsthm}
\usepackage{amsfonts}
\usepackage{amsmath}
\usepackage{bbm}
\usepackage{float}
\usepackage{url}
\usepackage{hyperref}
\usepackage[margin=0.5cm,font=footnotesize]{caption}
\usepackage{lineno}
\usepackage[round]{natbib}
\usepackage{appendix}
\usepackage{subfig}

\begin{document}
\title{\bf Development of a Thermodynamics of Human Cognition and Human Culture}
\author{Diederik Aerts, Jonito Aerts Arg\"uelles, Lester Beltran\footnote{Center Leo Apostel for Interdisciplinary Studies, 
        Free University of Brussels (VUB), Krijgskundestraat 33,
         1160 Brussels, Belgium; email addresses: diraerts@vub.be,jonitoarguelles@gmail.com,lestercc21@yahoo.com} 
        $\,$ and $\,$  Sandro Sozzo\footnote{Department of Humanities and Cultural Heritage (DIUM) and Centre CQSCS, University of Udine, Vicolo Florio 2/b, 33100 Udine, Italy; email address: sandro.sozzo@uniud.it}              }

\date{}

\maketitle
\begin{abstract}
\noindent
Inspired by foundational studies in classical and quantum physics, and by information retrieval studies in quantum information theory, we prove that the notions of `energy' and `entropy' can be consistently introduced in human language and, more generally, in human culture.
More explicitly, if energy is attributed to words according to their frequency of appearance in a text, then the ensuing energy levels are distributed non-classically, namely, they obey Bose-Einstein, rather than Maxwell-Boltzmann, statistics, as a consequence of the genuinely `quantum indistinguishability' of the words that appear in the text. Secondly, the `quantum entanglement' due to the  way meaning is carried by a text reduces the (von Neumann) entropy of the words that appear in the text, a behaviour which cannot be explained within classical (thermodynamic or information) entropy. We claim here that this `quantum-type behaviour is valid in general in human language', namely, any text is conceptually more concrete than the words composing it, which entails that the entropy of the overall text decreases. In addition, we provide examples taken from cognition, where quantization of energy appears in categorical perception, and from culture, where entities collaborate, thus `entangle', to decrease overall entropy. We use these findings to propose the development of a new `non-classical thermodynamic theory' for human cognition, which also covers broad parts of human culture and its artefacts and bridges concepts with quantum physics entities.
\end{abstract}

\section{Introduction \label{intro}} 
One of the most interesting aspects of human cognition concerns how people combine and exchange concepts through language and how new entities of meaning are formed through natural language. These processes are also important in applied cognition, e.g., computational linguistics, natural language processing and information retrieval where, in particular, the meaning content of words (thus, the corresponding concepts) and texts, is recovered through digital instruments.\footnote{Before proceeding further, it is important to clarify the terminology we use in this article. We use words, in particular the words that appear in written texts, as labels for concepts. Hence, the latter attribute meaning to the former. In our theoretical perspective, we consider concepts as abstract entities which are in a given state and whose state captures the meaning content of the concepts. As such, an individual concept is an example of an entity of meaning, where meaning is carried by the state which the concept is in. But, other entities of meaning can be constructed, e.g., conceptual combinations, which are concepts in itself, are also entities of meaning constructed by composing individual concepts. More generally, a piece of written text is an entity of meaning constructed by combining the concepts that correspond to the words appearing in the text. We will use this terminology consistently across the article.\label{wordconcept}}

Since 1990s, growing empirical evidence has convincingly demonstrated that the subtle and complex processes underlying human cognition, involving perception, language, judgements and decisions, cannot be represented in the mathematical formalisms originally conceived to deal with classical macroscopic systems (or entities), e.g., Boolean algebras, Kolmogorovian probabilities, commutative algebraic structures, fuzzy sets, and complexity theory (see, e.g., \citet{pitowsky1989}). On the other side, the use of the mathematical formalism of quantum theory in Hilbert space has been successful as a modelling, predictive and explanatory framework to represent entities in cognitive domains (see, e.g.,   \citet{aertsaerts1995,vanrijsbergen2004,aerts2009a,pothosbusemeyer2009,khrennikov2010,busemeyerbruza2012,aertsbroekaertgaborasozzo2013,aertsgaborasozzo2013,havenkhrennikov2013,kwampleskacbusemeyer2015,dallachiaragiuntininegri2015a,dallachiaragiuntininegri2015b,melucci2015,aertssozzoveloz2016,blutnerbeimgraben2016,broekaertetal2017,aertsetal2018} and references therein). The reason is that quantum structures, better than classical structures, are able to cope with the fundamental features of such entities, as `unavoidable uncertainty', `contextuality', `emergence', `indeterminism' and `superposition', which are thus not 
only characteristic and peculiar of microscopic entities, as originally believed in the early days of quantum theory.

Our research team has dedicated two decades to the investigation of the epistemological and mathematical differences between classical and quantum theories, firstly in physical and then in cognitive domains. This research has led to the development of a new `theoretical perspective' for human cognition in which words and their meaning content, hence concepts, are regarded as `entities that can be in different states and interact with each other and with contexts in a non-deterministic way'. In this perspective, words and concepts behave as quantum entities, as electrons, photons, etc., which justifies the use in cognition of the quantum formalism in Hilbert space, e.g., unit vectors, self-adjoint operators, unitary dynamics, probabilistic Born rule and tensor products (see, e.g., \citet{aertssozzo2016,aertssassolisozzo2016,pisanosozzo2020,aertssassolisozzoveloz2021}). On the other side, the analogies in physical and cognitive realms are so deep that we have also developed a new interpretation of quantum theory in which quantum entities themselves do not behave as physical objects but, rather, as concepts \citep{aerts2009b,aertsetal2020}.

Coming to the main content of this article, classical thermodynamics, as a physical theory, relies on the notions of `energy', whose conservation substantially expresses the content of its first law, and `entropy', which is the main variable in one of the formulations of the second law. Thermodynamic entropy can then be expressed, in a formally analogous way, within classical information theory, in terms of `Shannon (information) entropy' \citep{shannon1948a,shannon1948b}. Finally, classical statistical mechanics provides a general theoretical framework where thermodynamic theory can be properly placed. We intend to use some recent results, which extend our `quantum cognition theoretical perspective to a large number of conceptual entities' \citep{aertssozzoveloz2015,aertsbeltran2020,beltran2021,aertsbeltran2022a,aertsbeltran2022b}, and propose the new idea that a thermodynamic theory can also be elaborated for human language in the first place and, more generally, for human cognition and for entities, such as human artefacts, within human culture, which relies on the notions of energy and entropy, once these are consistently defined in the cognitive realm and in the realm of human culture. However, we also demonstrate that it has to be a `non-classical thermodynamic theory', because energy and entropy have genuinely quantum features in these realms. Before proceeding further, it is worth to summarise the recent results above, which are reviewed and put into a new and unified theoretical perspective in Sections \ref{energy} and \ref{entropy}.

Firstly, we have proved through various empirical examples, that, if one considers a large text, which can be treated as a combination of several words, then this behaves at a statistical level as a `quantum gas', specifically, a `gas of bosons in thermal equilibrium with its environment in a state that is close to a Bose-Einstein condensate near absolute zero temperature' \citep{aertsbeltran2020,beltran2021,aertsbeltran2022a,aertsbeltran2022b}. To this end, we have attributed `energy levels' to words, in an inverse proportion to their frequency of appearance in the text. Then, we have proved that the distribution of these energy levels across words does not obey `Maxwell-Boltzmann statistics', which would hold for identical and distinguishable entities, but it obeys `Bose-Einstein statistics', which would hold for identical and undistinguishable entities. The reason for the appearance of Bose-Einstein statistics is due to the fact that identical words can be exchanged in position within the text without changing the text itself, hence they behave indeed as `indistinguishable', exactly as bosons, e.g., photons, whose wave functions are symmetric with respect to the exchange of boson positions \citep{huang1987}. In addition, the `meaning dynamics' between the words appearing in a text make most frequent, thus lowest energy, words more likely to appear in the text than less frequent, and we showed on concrete texts that this happens obeying a specific Bose-Einstein statistical distribution characterizing a behaviour  
that closely resembles a Bose-Einstein condensate close to absolute zero temperature \citep{cornellwieman2002,ketterle2002}. To make the analogy with quantum physics more explicit and convincing, we have introduced the notion of `cogniton' as the fundamental quantum of human cognition, such that a text behaves as a `Bose-Einstein gas of cognitons' \citep{aertsbeltran2020,aertsbeltran2022a,aertsbeltran2022b}. In addition, we have provided a theoretical foundation of `Zipf's law' in language \citep{zipf1935,zipf1949}, whose theoretical origin is not understood yet and whose nature is therefore considered to be empirical, while our results \citep{aertsbeltran2020,aertsbeltran2022a,aertsbeltran2022b} derive it on the basis of 
the presence of a Bose-Einstein behaviour of energy levels of words in a text.

Secondly, we have proved in various empirical tests on human participants and also in information retrieval tests on the web, that various combinations of two concepts systematically violate `Bell's inequalities' \citep{aertssozzo2011,aertssozzo2014,beltrangeriente2018,aertsetal2019,arguellessozzo2020,aertsetal2021,aertsetal2023a}, a fact that is considered in physics as a consequence of the presence of `quantum entanglement'  \citep{bell1964,clauseretal1969,aspectdalibardroger1982,vienna2013,urbana2013}. In the case of conceptual entities, this entanglement is produced by the connections in meaning that are created whenever two or more concepts combine to form more complex entities. These connections give rise to non-classical statistical correlations, exactly as occurs with microscopic quantum entities, as electrons, photons, etc. In addition, we have calculated the `von Neumann entropy' of the component concepts contained in entangled combinations of two concepts,
    finding that the entropy is systematically greater than zero \citep{aertsbeltran2022a}. Since in a pure entangled state representing the combined 
    concept, the von Neumann entropy is zero, the identified `sub-additivity of entropy' indicates that the component 
    concepts 
    are in a non-pure state (or `density state') \citep{nielsenchuang2000}. Hence, a `reduction of entropy' occurs whenever two 
    concepts    combine to form a composed conceptual entity, which is completely analogous to what occurs in the composition of quantum physics entities. This strongly indicates that a notion of entropy can be consistently introduced for simple texts consisting of a few words which behaves in a way that substantially differs from (thermodynamic or information) entropy, as the entropy of classical entities generally increases as a consequence of composition.

We firstly claim that the non-classical result above that we have identified on simple texts made up of two words (see the connection between words, concepts and entities of meaning illustrated in footnote \ref{wordconcept}) is absolutely general in human 
language, in the sense that any text expressing a meaning content is systematically more concrete, at a conceptual level, than the words composing it. This means that one can naturally assume that the meaning content of the overall text can be represented by a `zero-entropy pure entangled state', whereas the component words are meaning-wise represented by `positive-entropy non-pure (density) states'. Indeed, words, carrying in general quite abstract meaning, in a contextual way, the context being primarily the surrounding words, sentences, but possibly the whole text, become contextual carriers of more concrete meaning, and entanglement and how the von Neumann entropy behaves is structurally and mathematically expressing this phenomenon.

Secondly, putting the results above into a unified perspective, we suggest that energy behaves non-classically in other areas of human cognition, not only language. For example, the typical `warping phenomenon in categorical perception', which is omnipresent in cognition and where groups of stimuli clump together, can be considered as a `phenomenon of quantization', where individual concepts play the role of the quanta of cognition (see \citet{aertsaertsarguelles2022}). We have not investigated yet in a general way the statistical behaviour of the quantization that is engendered by this phenomenon of categorical perception and its resulting quanta, but, since human language can be considered to be an example of it, and all properties that made us derive the Bose-Einstein statistics for human language are also satisfied within it, we do not expect a different structure than the Bose-Einstein behaviour identified in language.

Thirdly, we believe that entanglement is also present in those realms of culture where human artefacts are involved. For example, the `hunting activity', where different individuals collaborate to reduce the uncertainty of the whole and make a successful hunting, can be considered as a process of entanglement of the parts which reduces the entropy of the whole \citep{aertsetal2023b}. This is why we believe that the mathematical theory formalising the behaviour of conceptual entities, in every domain in which these appear, be it language, cognition or culture, has to be a `quantum-type thermodynamics'. These results are explored in Section \ref{thermodynamics}, where we also provide reasonable arguments to claim that the direction indicated here for a quantum-type thermodynamics of cognition and culture can also further, certainly with respect to understanding what is ongoing, the development of a `quantum thermodynamics for physics', which is still highly unfinished conceptually \citep{gemmeretal2009,mahler2015,gogolineisert2016}.

\section{Energy and Bose-Einstein statistics in large texts\label{energy}}
In this section, we analyse and deepen our recent findings on the identification of Bose-Einstein statistics in large texts, and put these results in the perspective of a quantum-type thermodynamic theory of human cognition.

It is well know in physics that quantum entities with integer spin, also called `bosons', e.g., photons or light particles, are characterised by wave functions that are symmetric with respect to boson positions exchange. Then, the indistinguishable nature of identical bosons determines a non-classical statistical behaviour, i.e. different from `Maxwell-Boltzmann statistics', whenever a large number of these quantum entities are considered overall \citep{huang1987}. For a dilute gas of bosons close to absolute zero temperature in thermal equilibrium with environment, all bosons tend to occupy the lowest energy state, a phenomenon known as `Bose-Einstein condensation' and well confirmed empirically \citep{cornellwieman2002,ketterle2002}. 

As sketched in Section \ref{intro}, we have recently identified a `completely analogous statistical behaviour' in human language,
namely, if one considers a large story-telling text and attributes energy levels to the words appearing in the text, according to their numbers of appearance, then the component words exhibit `genuinely quantum aspects', namely, superposition, entanglement and overlapping de Broglie wave functions. As a consequence, the overall entity of meaning made up of the concepts associated with the words that appear in the text
satisfies the statistical properties of a `Bose-Einstein condensate' \citep{aertsbeltran2020,aertsbeltran2022a,aertsbeltran2022b}. 

We do not report the full technical details of these studies in this article. In this regard, we refer the interested reader to \citet{aertssozzoveloz2015,aertsbeltran2020,beltran2021,aertsbeltran2022a,aertsbeltran2022b} and references therein. We instead would like to show that these results support the claim that, once the notion of `energy' is properly introduced in cognitive domains, this variable systematically behaves in a non-classical way, i.e. giving rise to a Bose-Einstein statistical distribution and not to a Maxwell-Boltzmann one. To this end, let us consider a story-telling text, together with its content in terms of meaning. For example, we have analysed in detail the Winnie the Pooh story entitled ``In Which Piglet Meets a Haffalump'' \citep{milne1926} in \citet{aertsbeltran2020}. However, we stress that the analysis has been repeated with several texts, including short and long stories, e.g., novels, and we always found the results presented in the following.

Let us see how energy can be appropriately quantified in the linguistic realm. Each word can be associated with a conceptual entity in a specific state (see Footnote \ref{wordconcept}) whose energy level is defined by the number of times the word appears in the given text. More precisely, the most frequent word is given the lowest energy level $E_0$. Now, suppose the text contains $n+1$ different words. Let us then order them according to the increasing energy level or, equivalently, according to their decreasing order of appearance in the text, and set, for a given word $w_i$, $E_i=i$, $i=0,1,\ldots,n$ (hence, we set $E_0=0$ as the `ground state energy').

Next, let $N(E_i)$ be the number of times the word $w_i$, with energy $E_i$, appears in the text. Hence, the `total number of words' appearing in the text is
\begin{equation} \label{totalnumber}
N=\sum_{i=0}^{n} N(E_i)
\end{equation}
It follows from the above that the word $w_i$ is associated with the energy $E_i N(E_i)$, $i=0,1,\ldots,n$, hence, the `total energy of words' is
\begin{equation} \label{totalenergy}
E=\sum_{i=0}^{n} E_i N(E_i)=\sum_{i=0}^{n} i N(E_i)
\end{equation}
As in physics, both $N$ and $E$ can be retrieved from empirical data, word counts in the case of a text.

Before proceeding further, it is important to stress a difference between physics and cognition with respect to measures of energy: in physics, the unit of energy is a derived quantity, which is measured in, e.g., $Kg\cdot m^2\cdot s^{-2}$ in the International System of Units. On the contrary, energy is a fundamental quantity in cognition, where the cognitive equivalent of physical space cannot be uniquely identified.

Shall we expect the numbers $N(E_i)$ in Equations (\ref{totalnumber}) and (\ref{totalenergy}) to satisfy any particular statistical distribution? To understand this important point, let us consider, e.g., the concept combination {\it Eleven Animals}. It is clear that, at a conceptual level, each one of the eleven animals is completely `identical with' and `indistinguishable from' each other of the eleven animals. On the other side, it is likewise clear that, in the case of `eleven physical animals', there are always differences between each one of the eleven animals, because as `objects' present in the physical world, they have an individuality and, as individuals with spatially localized physical bodies, none of them is really identical to the others. This means that each one of the animals can always be distinguished from the others. In the latter case, we expect Maxwell-Boltzmann statistics to apply as a consequence of the fact that, due to their physical nature, the eleven animals are not completely identical, hence distinguishable. In the former case, we instead expect a non-classical statistics to apply as a consequence of the fact that, due to their conceptual nature, the eleven animals are identical, hence intrinsically indistinguishable.

But, then why, shall we expect that exactly Bose-Einstein statistics works in the case of words (concepts) in a large text? Again, an example can help us to grasp the point. Let us consider any text and, e.g., two instances of the word (concept) {\it Cat} appearing in the text. Then, if we exchange one of the words (concepts) {\it Cat} with the other word (concept) {\it Cat}, absolutely nothing changes in the text. Thus, a text contains a perfect symmetry for the exchange of words (concepts, or also `cognitons', introducing what we have defined as the `quanta of human cognition') in the same state, exactly as in the case of identical and indistinguishable bosons.

Let us reinforce the point above and illustrate in a simple situation the fundamental differences between the two distributions, Bose-Einstein and Maxwell-Boltzmann, in terms of `statistical dependence'. More precisely, consider the example of two entities $S_1$ and $S_2$ which can be in two different states $p_1$ and $p_2$, respectively. If we analyse this example from a Maxwell-Boltzmann point of view, four different configurations may occur: (i) $S_1$ and $S_2$ are both in the state $p_1$, (ii) $S_1$ and $S_2$ are both in the state $p_2$, (iii) $S_1$ is in the state $p_1$ and $S_2$ is in the state $p_2$, and (iv) $S_1$ is in the state $p_2$ and $S_2$ is in the state $p_1$. Each of these configurations occur with probability 1/4, as a consequence of statistical independence and distinguishability. If we instead analyse the example from a Bose-Einstein point of view, only three configurations may occur: (i) $S_1$ and $S_2$ are both in the state $p_1$, (ii) $S_1$ and $S_2$ are both in the state $p_2$, (iii) one entity is in the state $p_1$ and the other is in the state $p_2$. Each of these configurations occur with probability 1/3, and this is incompatible with an independent behaviour of the individual entities. Even assuming an `epistemic-only indistinguishability' of (iii) and (iv) in the Maxwell-Boltzmann case, one would then get three configurations with probabilities 1/4, 1/4, 1/2, which is again different from 1/3, 1/3, 1/3. By comparing (i), (ii) and (iii) in the two distributions, we notice that Bose--Einstein statistics assigns `a higher probability' to configurations (i) and (ii), i.e. both $S_1$ and $S_2$ in the state $p_1$ and both $S_1$ and $S_2$ are in the state $p_2$, respectively. In other words, Bose-Einstein statistics predicts that `entities bundle together in the same state, more than one would expect'. A Bose-Einstein condensate reflects this behaviour of `bundling together in the same state', the state of lowest energy in this case.

Let us now illustrate the counterpart of situations (i)--(iii) in the conceptual realm. To this end, consider the concepts {\it Cat} and {\it Dog} and the configurations {\it Two Cats}, {\it Two Dogs} and {\it One Cat And One Dog}. Both {\it Cat} and {\it Dog} can be regarded as states of the concept {\it Animal}, hence we are studying the situation in which two entities {\it Animal} can be in two different states {\it Cat} and {\it Dog}. We firstly consider a Maxwell-Boltzmann situation with respect to these three configurations. Suppose we visit a farm with a lot of animals, all of them being cats or dogs and living in the farm, more or less in equal number, and we receive as a present two of them randomly chosen by the farmer. Then, the chance that the gift will be one cat and one dog will be twice that of it being either two cats or two dogs. Hence, the probability is 1/4 for two cats, 1/4 for two dogs, and 1/2 for one cat and one dog. Indeed, we do not see any difference between `one cat and one dog' or `one dog and one cat'. Even so, the situation that we have called `epistemic-only indistinguishability' in Maxwell-Boltzmann statistics gives rise to the probabilities 1/4, 1/4, 1/2, not to the Bose-Einstein probabilities 1/3, 1/3, 1/3. What, however, if we asked a child who has been promised he/she can have two pets and choose for himself/herself whether either pet is a cat or a dog. The micro-states that come into play in this case exist in the realm of the child's conceptual world, and there is no reason that within this conceptual world there will be a double amount of (micro-)states for the choice of a cat and a dog as compared to the choices for two cats or two dogs, which leads to 1/3, 1/3, 1/3 for the probabilities, hence to Bose-Einstein statistics. Also, the micro-states in the child's conceptual world will no longer be states of, on the one hand, cat, separately, and on the other hand, dog, separately, but will be micro-states of the three combinations themselves, two cats, two dogs, or, one cat and one dog. Indeed, in the child's conceptual world, either there are two kitties always playing together, two puppies always playing together, or a kittie and a puppy playing together. \footnote{At first glance, there seems to be no reason why the child will exactly form 1/3, 1/3, and 1/3 as probabilities with respect to the micro-states. For example, it could be that `a cat and a dog' is even less likely to be chosen 
%%D that 
than the 1/3 that Bose-Einstein statistics assigns to it (e.g., the child may have been told that a cat and a dog could lead to problems later, when they grow up). However, we should remember in this regard that we are talking about `statistics' and, if one considers all situations of children who will choose, and similar situations with different choices then, for symmetry reasons, the 1/3, 1/3, 1/3 will probably result for the statistics.} In the child's feeling and thinking about his/her two pets, there is no functionality to assign, separate (micro-)states to each of the pets separately, as there must be if a Maxwell-Boltzmann statistics is to be obtained. The bodies of the pets, since they are macroscopic material entities, always remain distinct, which is why a classical mechanics description of these bodies will indeed lead to a Maxwell-Boltzmann statistics. For those who have experience with the Hilbert space formalism of quantum theory, what the child is doing in its conceptual realm with the two pets will seem familiar, indeed, when we describe two quantum entities by the tensor product of the two Hilbert spaces describing each of the component entities, then it is exactly `that' we are doing. The vast majority of (micro-)states in the tensor product are directly states of the two entities together, and not reducible to states of each of them separately, these are the product states, but it is a minority. We call `entangled states' these states not reducible to product states. The similarity can perhaps be even better noted thinking of the Schr\"odinger wave quantum description, where the wave function has variables in configuration space. As a result, the wave function of two quantum entities is a function of six positional variables, which in general, if there is entanglement, cannot be written as the product of two wave functions of three positional variables. As if the measuring apparatus in quantum theory uses a kind of conceptual space for the interaction with the quantum entity to take place \citep{aertsaertsarguelles2022}.

Given the above analysis, we can easily imagine that, for a text with meaning, the micro-states are mostly assigned to the parts of the text that carry the meaning, often even to the whole text, and that the micro-states assigned to individual words in the text -- that is, the product states -- are in the minority, since these individual words, as a consequence of entanglement and contextuality, have often surrendered their individuality at the expense of the purity (of state) of the whole text, as a function of its carrying capacity with respect to `meaning'.

Coming to the mathematical representation, Bose-Einstein statistics would entail the $N(E_i)$s to satisfy 
\begin{equation} \label{be}
N(E_i)=\frac{1}{Ae^{E_i/B}-1}
\end{equation}
where $A$ and $B$ are constants to be determined by imposing Equations (\ref{totalnumber}) and (\ref{totalenergy}) to hold. On the contrary, Maxwell-Boltzmann statistics would entail the $N(E_i)$s to satisfy 
\begin{equation} \label{mb}
N(E_i)=\frac{1}{Ce^{E_i/D}}
\end{equation}
where $C$ and $D$ are again constants to be determined from Equations (\ref{totalnumber}) and (\ref{totalenergy}).

\begin{figure}
    \centering
    \subfloat[Numbers of appearances distribution graphs]{{\includegraphics[scale=0.42]{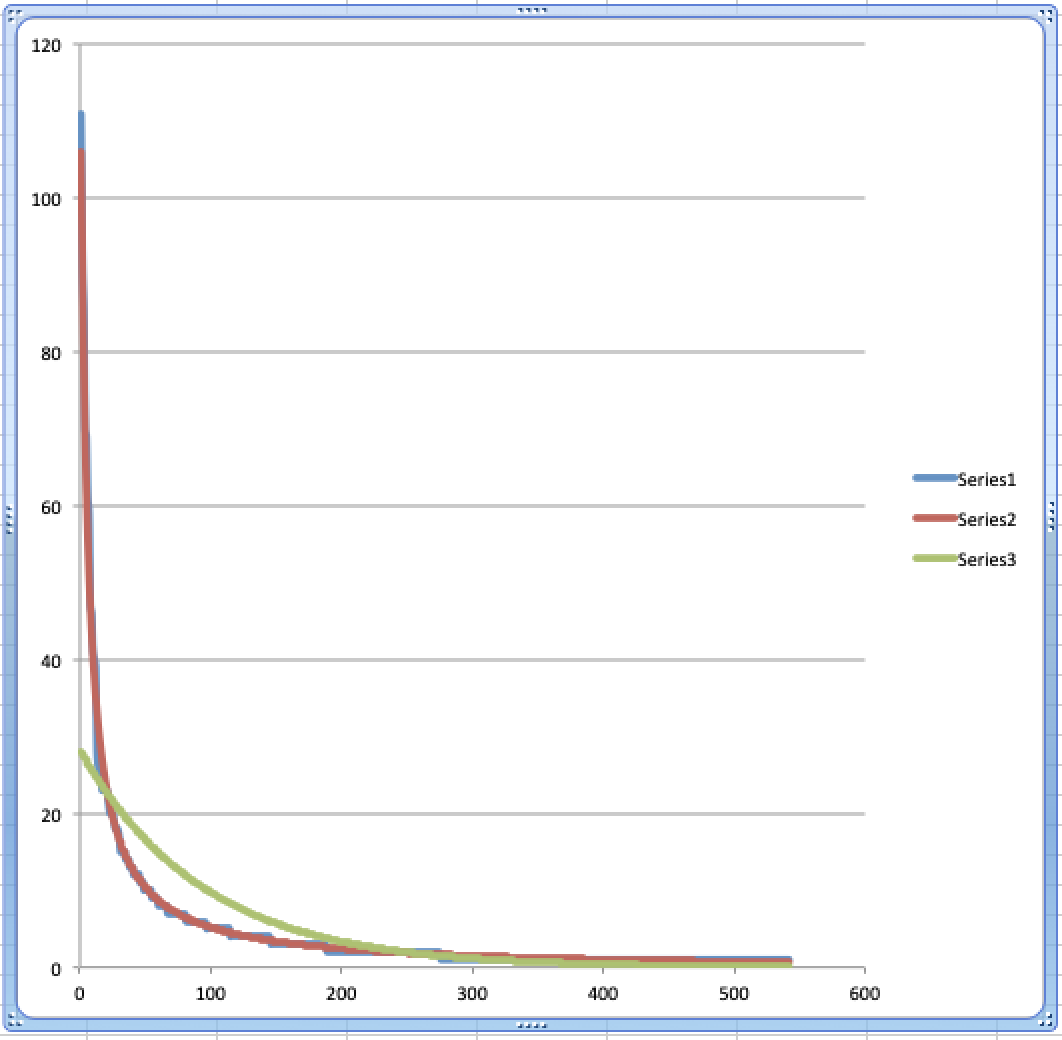} }}%
    \qquad
    \subfloat[$\log/\log$ graphs of numbers of appearances distributions]{{\includegraphics[scale=0.42]{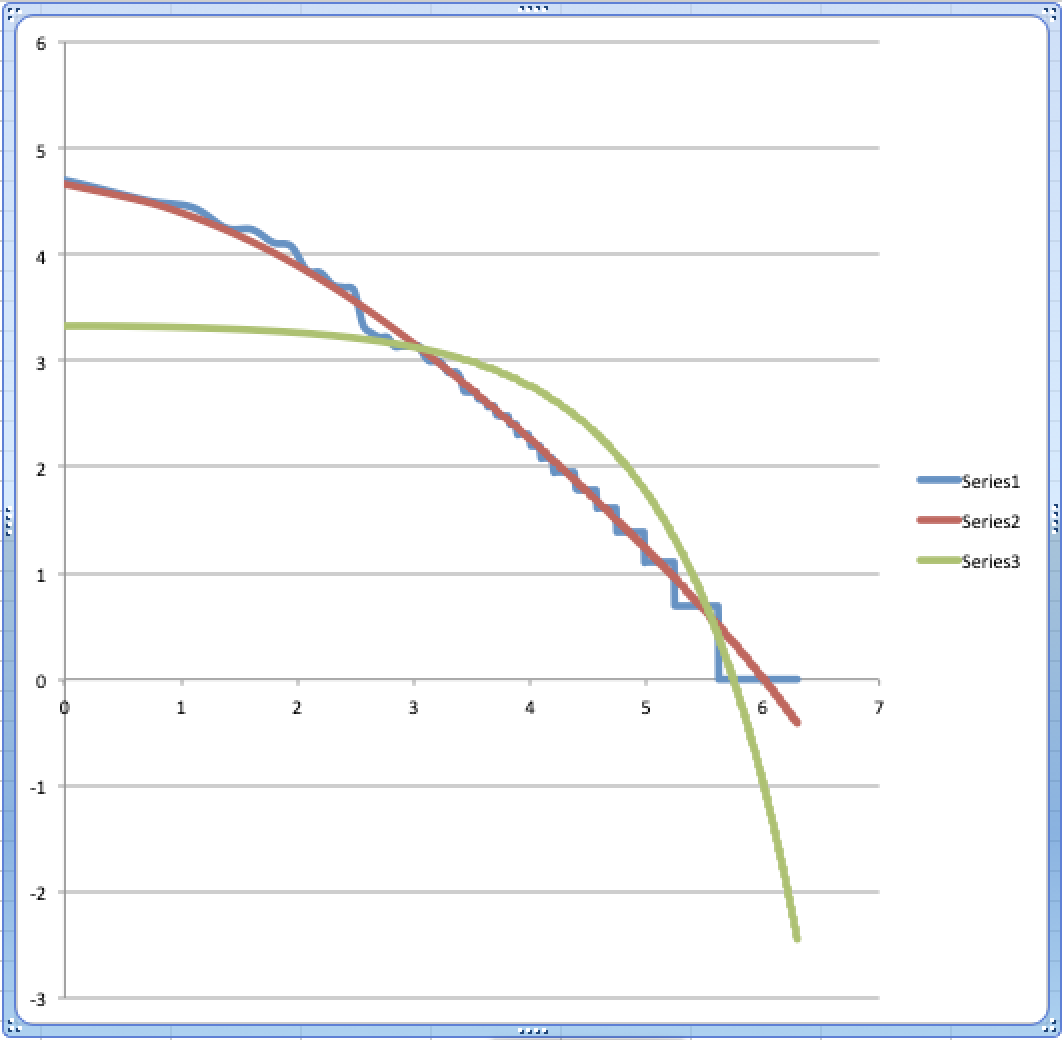} }}%
    \caption{In (a) we represent the number of appearances of words in the Winnie the Pooh story `In Which Piglet Meets a Haffalump' \citep{milne1926}, ranked from lowest energy level, corresponding to the most often appearing word, to highest energy level, corresponding to the least often appearing word. The blue graph (Series 1) represents the data, i.e. the collected numbers of appearances from the story, the red graph (Series 2) is a Bose-Einstein distribution model for these numbers of appearances, and the green graph (Series 3) is a Maxwell-Boltzmann distribution model. In (b) we represent the $\log / \log$ graphs of the numbers of appearances and their Bose--Einstein and Maxwell-Boltzmann models. The red and blue graphs coincide almost completely in both (a) and (b), whereas the green graph does not coincide at all with the blue graph of the data. This shows that the Bose-Einstein statistical distribution is a good model for the numbers of appearances, while the Maxwell-Boltzmann distribution is not.}%
    \label{piglethaffalunmpgraphpiglethaffalunmploggraph}%
\end{figure}

\begin{figure}
\centering
\includegraphics[scale=0.42]{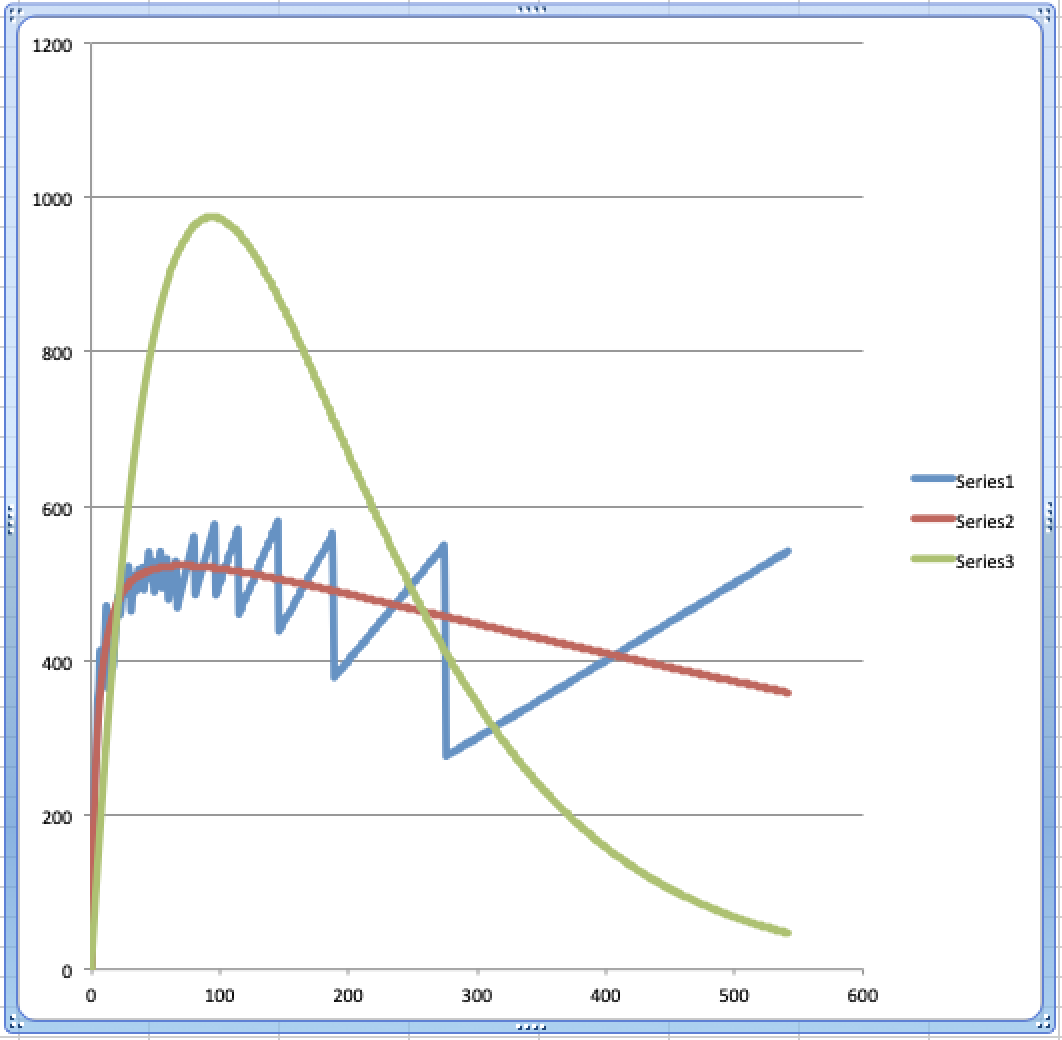}
\caption{A representation of the energy distribution of the Winnie the Pooh story `In Which Piglet Meets a Haffalump' \citep{milne1926}. The blue graph (Series 1) represents the energy radiated by the story per energy level, the red graph (Series~2) represents the energy radiated by the Bose-Einstein model of the story per energy level, and the green graph (Series 3) represents the energy radiated by the Maxwell-Boltzmann model of the story per energy level.}
\label{piglethaffalunmpenergygraph}
\end{figure}

In \citet{aertsbeltran2020,aertsbeltran2022a,aertsbeltran2022b}, we have determined the constants $A$ and $B$ in Equation (\ref{be}) and the constants $C$ and $D$ in Equation (\ref{mb}), and compared the ensuing Bose-Einstein and Maxwell-Boltzmann distributions with empirical data. Figures \ref{piglethaffalunmpgraphpiglethaffalunmploggraph} and \ref{piglethaffalunmpenergygraph} summarise the comparison. 
The results of the comparison significantly show that `Bose-Einstein statistics is in remarkably good fit with empirical data', whereas a `large deviation is observed in the data from Maxwell-Boltzmann statistics' \citep{aertsbeltran2020,aertsbeltran2022a}.

The conclusion is that, once a notion of energy is adequately introduced and quantified from empirical data, a collection of words in a story-telling text behaves as a suitable gas of bosons. As a matter of fact, we have introduced the term `cogniton' as the fundamental quantum of cognition. Equivalently, each word $w_i$ with energy $E_i$ corresponds to a state of a cogniton with energy $E_i$. The overall text then behaves as a `gas of cognitons' whose energies are distributed according to Bose-Einstein statistics.
 
Furthermore,
it should be noted that the analogy between quantum physics and 
language is even deeper and more impressive. In this regard, we have anticipated at the beginning of this section that a dilute gas of bosons, in thermal equilibrium with environment, tends to behave as a Bose-Einstein condensate when temperature gets close to absolute zero temperature. In this case, indeed, the available energy is so little that all boson particles are `forced' to make a transition to their `lowest energy state'. Under these conditions of `quantum coherence', physicists typically say that the `de Broglie wave functions' start to overlap and the gas behaves in a genuinely quantum way \citep{huang1987}. 

In \citet{aertsbeltran2022a,aertsbeltran2022b}, we have demonstrated that the presence of meaning makes in language the same effect that this quantum coherence makes in physics, namely, a gas of cognitons behaves as a dilute Bose-Einstein condensate in which most of the cognitons tend to occupy the lowest energy state. Let us note that a text telling a story in that sense is not at the temperature where the full Bose-Einstein condensate realizes but, rather, at a temperature close to it, where there is still enough variety of energy levels to allow the necessary meaningful words to form the story, through the cogniton that can nestle in those available energy levels, and thus form the different words needed for the story. However, the temperature must be low enough for all cognitons involved to be fundamentally entangled, because only 
by means of this entanglement can the words of a story carry meaning.

Before concluding the section, it is worth mentioning another interesting result that we have found within our new theoretical perspective: the attribution of specific energies to words in large texts also provides a theoretical foundation to a law of language, whose nature is generally considered as empirical and which is studied abundantly in the literature, namely, `Zipf's law' \citep{zipf1935,zipf1949}. Indeed, when ranking words according to their number of appearance in a given text, one discovers that, in a text with $n$ words, the rank $R_i$ and the number of appearance $N_i$ of the word $w_i$, satisfy the equation
\begin{equation}\label{zipf}
R_iN_i={\textrm{constant}}
\end{equation}
In \citet{aertsbeltran2020}, we have provided theoretical arguments to show that Zipf's law can be rigorously derived from the Bose-Einstein statistics underlying words within a given text.\footnote{It is worth mentioning two aspects of Zipf's law. Firstly, while the law was discovered in quantitative linguistics, it also systematically appears in a variety of human-created systems, such as rankings of the size of cities \citep{gabaix1999}, rankings of the size of income \citep{aokimakoto2017}, but also rankings that seem much more marginal, such as the number of people who watch the same TV station \citep{erikssonetal2013}, or the rankings of notes in music \citep{zanette2006}, or the rankings of cells’ transcriptomes \citep{lazzardietal2021}. The theoretical foundation that follows from our research characterizes Zipf's law in human language as the equivalent of Planck's law for the photons of light (see \citet{aertsbeltran2022b} where this equivalence is used to propose a radiation law for human communication through human language). Our theoretical foundation also explains why Zipf's law can be found in so many other areas of human culture where rankings can be discerned. Indeed, according to our analysis, they are rankings of different energy levels; in other words, with our derivation of Zipf's law, the rankings identified become the equivalent of the spectra found in physical material entities. Hence, our explanation here supports the claim in Section \ref{thermodynamics} that a quantum-type thermodynamics can be constructed for human culture, not only for language or cognition, as a consequence of Zipf's law being so abundant in human culture. Secondly, according to some authors \citep{perline1996}, the law can be derived from the central limit theorem. At the best of our knowledge, the literature does not converge on this conclusion (see, e.g., \citet{trollbeimgraben1998}). On the other side, it is generally accepted that the origins of this law are not completely understood. In our opinion, we have provided a robust explanation for its appearance in language and culture.}

We believe that the results presented in this section have a general validity in language. More precisely, the meaning content of the concepts associated with the words that appear in any text exhibit genuinely quantum features, that is, quantization, coherence, Bose-Einstein indistinguishability, and condensation. These features arise whenever one suitably associates energy levels to words and determine the energy of the overall text to behave `macroscopically' in a non-classical way. Hence, a thermodynamic theory of language which would incorporate energy cannot be classical thermodynamics, because it should take into account these non-classical, genuinely quantum-type, features of energy. 
However, we also believe that the results proposed in this section can be extended to human cognition in general, and more so apply not only to words of human language, but also to general entities belonging to human culture, such as artefacts. A first important piece of evidence is the occurrence of Zipf's law in broad layers of human culture, and the equivalence of the thermodynamic energy scheme we propose with the ranking structure used in Zipf's law. In the remainder of this article, we will present two other pieces of evidence, one that relies on the phenomenon of categorical perception and its omnipresence in human cognition, and the second with how macroscopic material entities can easily be used to realize entanglement. 
A development of a quantum-type thermodynamics in these domains would also be beneficial to the further development of a thermodynamics of quantum physics, which struggles with several issues of understanding \citep{gemmeretal2009,mahler2015,gogolineisert2016}. 

In the next section, we intend to show that a notion of entropy, if suitably introduced in language, exhibits similar non-classical features.

\section{Entanglement and entropy in the combination of concepts\label{entropy}}
In this section, we analyse and deepen our recent findings on the identification of entanglement in the combination of words (concepts) and put these results in the perspective of a quantum-type thermodynamic theory of human cognition.

It is well know in physics that `Bell's inequalities' provide a technical constraint to represent the statistical correlations of composite entities within classical probabilistic, i.e. Kolmogorovian, formalisms (see, e.g., \citet{pitowsky1989}). Then, the violation of Bell's inequalities in quantum theory, which have been largely confirmed empirically \citep{aspectdalibardroger1982,vienna2013,urbana2013}, proves that entities as photons and electrons exhibit genuinely non-classical features, namely, `contextuality', `nonlocality', `non-separability', and probably the most important one, `entanglement' \citep{bell1964,clauseretal1969}. Entanglement, in particular, entails that these entities can be connected in a composite, e.g., bipartite, entity so strongly that they kind of lose their identity.

A typical setting of a Bell-type test consists of a composite bipartite physical entity $S_{12}$, that is prepared in a given initial state and is such that two individual entities $S_1$ and $S_2$ can be recognised as parts of $S_{12}$. Then, four coincidence experiments $AB$, $AB'$, $A'B$ and $A'B'$ are performed on $S_{12}$ which consist in performing experiments $A$ with outcomes $A_1$ and $A_2$ and $A'$ with outcomes $A'_1$ and $A'_2$ on $S_1$, and experiments $B$ with outcomes $B_1$ and $B_2$ and $B'$ with outcomes $B'_1$ and $B'_2$ on $S_2$. If the experiment outcomes can only be $-1$ or $+1$, then the expected values of $AB$, $AB'$, $A'B$ and $A'B'$ become the `correlation functions' $E(AB)$, $E(AB')$, $E(A'B)$ and $E(A'B')$, respectively. In this case, the correlation functions can be represented within a classical probabilistic formalism if and only if they satisfy the `Clauser-Horne-Shimony-Holt (CHSH) version' of Bell's inequalities \citep{clauseretal1969}, as follows:
\begin{equation} \label{chsh}
-2 \le \Delta_{CHSH} \le +2
\end{equation}
where $\Delta_{CHSH}$ is the `CHSH factor' which is defined as
\begin{equation} \label{CHSH_factor}
\Delta_{CHSH}=E(A'B')+E(A'B)+E(AB')-E(AB)
\end{equation}
On the other side, composite bipartite quantum entities can be prepared in `entangled states' such that Equation (\ref{chsh}) is violated. Hence, it is entanglement that is crucial for the violation of Bell's inequalities in quantum theory. The presence of entanglement can be revealed by explicitly calculating the `von Neumann entropy' associated with the component entities \citep{nielsenchuang2000}. Let us thus review the notion of von Neumann entropy and its relationship with entanglement, since these are important to attain our main results in this section.

For a given quantum entity in a state represented by the density operator $\rho$  on the Hilbert space $\mathscr H$,\footnote{We recall that a density operator $\rho$ is a positive trace-class self-adjoint operator on the Hilbert space $\mathscr H$ associated with the entity such that $\textrm{Tr}\rho=1$, where $\textrm{Tr}$ denotes the trace operation. Pure states are represented by particular cases of density operators $\rho=|\psi\rangle\langle\psi|$, where $|\psi\rangle$ is a unit vector of $\mathscr H$ (equivalently, one says that a pure state is represented by a unit vector $|\psi\rangle \in \mathscr H$). In the case of a pure state, one has $\textrm{Tr}\rho^2=\textrm{Tr}\rho=1$. In the case of a non-pure state, one instead has $\textrm{Tr}\rho^2<\textrm{Tr}\rho=1$. \label{staterepresentation}} the von Neumann entropy is defined as
\begin{equation} \label{vonnneumann}
S(\rho)=-\textrm{Tr}\rho \log_2 \rho
\end{equation}
If one considers the spectral decomposition of $\rho$, i.e. $\rho=\sum_{i=1}^{n}p_i |\psi_i\rangle\langle\psi_i|$, where $p_i\ge 0$, $i=1,\ldots,n$, $\sum_{i=1}^{n}p_i=1$ and $\{|\psi_i\rangle \}_{i=1,\ldots,n}$ is an orthonormal (ON) basis in $\mathscr H$, then Equation (\ref{vonnneumann}) can be rewritten as
\begin{equation} \label{shannon}
S(\rho)=-\sum_{i=1}^{n} p_i \log_2 p_i
\end{equation}
which coincides with `Shannon entropy'\footnote{Shannon entropy is typically considered as the information-theoretic counterpart of the entropy of a classical thermodynamic entity \citep{shannon1948a,shannon1948b}.} for a classical probability distribution generated by the $p_i$s. One can easily prove that $S(\rho) \ge 0$, and $S(\rho)=0$ if and only if $\rho$ represents a pure state, that is, $\rho=|\psi\rangle\langle\psi|$, for some unit vector $|\psi\rangle \in \mathscr H$ \citep{nielsenchuang2000}. 

Suppose now that a composite bipartite entity $S_{12}$ is in a pure entangled state represented by the density operator $\rho_{12}=|\Psi_{12}\rangle\langle \Psi_{12}|$, where $|\Psi_{12}\rangle$ is a unit vector of the Hilbert space associated with $S_{12}$. We know that the von Neumann entropy is $S(\rho_{12})=0$. On the other side, the states of the component entities $S_1$ and $S_2$ are represented by density operators $\rho_1$ and $\rho_2$ obtained through the partial trace operation with respect to $S_2$ and $S_1$, respectively, that is, $\rho_1=\textrm{Tr}_2 \rho_{12}$ and $\rho_2=\textrm{Tr}_{1}\rho_{12}$. In this case, one easily proves that $\rho_1$ and $\rho_2$ do represent non-pure states, also called `density states' \citep{nielsenchuang2000}. In addition, using the Schmidt-von Neumann decomposition of $|\Psi_{12}\rangle$, one gets that $S(\rho_1)=S(\rho_2) >0$ (for example, a maximally entangled state in a 2-dimensional Hilbert space gives $S(\rho_1)=S(\rho_2)=\log_2 2$).\footnote{This result agrees with two general properties of von Neumann entropy, (i) $S(\rho_{12})\le S(\rho_1)+S(\rho_2)$, called `sub-additivity', and (ii) $S(\rho_{12}) \ge |S(\rho_1)-S(\rho_2)|$, sometimes called the `Araki-Lieb inequality' \citep{nielsenchuang2000}.} This behaviour of the entropy of a composite bipartite physical entity in quantum theory does not occur with the entropy of a classical thermodynamic entity and is generally considered as a `measure of entanglement' \citep{horodeckis2009}.

We intend to prove that, once a notion of entropy is introduced for a combination of concepts, this exactly behaves as the von Neumann entropy of a quantum entity. To this end, we need to put into perspective the empirical and theoretical studies we have  produced in regard to the violation of Bell's inequalities in cognitive domains \citep{aertssozzo2011,aertssozzo2014,beltrangeriente2018,aertsetal2019,arguellessozzo2020,aertsetal2021,aertsetal2023a}.

In the last decade, we have performed several empirical tests on Bell's inequalities and the possible presence of entanglement in cognition, using different combination of two concepts. In particular, we have performed two tests involving human participants \citep{aertssozzo2011,aertsetal2023a}, three documents retrieval tests on the web, using the corpuses of documents `Google Books', `Contemporary American English (COCA)' and `News on Web (NOW)' \citep{beltrangeriente2018,aertsetal2021}, and one image retrieval test using the search engine `Google Images' \citep{arguellessozzo2020}. In all these tests, we have used the concept combination {\it The Animal Acts} which we have considered as a composite bipartite conceptual entity made up of the component entities {\it Animal} and {\it Acts}, where the latter word refers to the sound, or noise, produced by an animal.

The collected data that are relevant to the purposes of this article are reported in Table \ref{animalactsentanglements}. As we can see, all data significantly violate Bell's inequalities by amounts that are very close to the violation in quantum physics tests.\footnote{It is well known that quantum theory predicts that $\Delta_{CHSH}=2\sqrt{2}\approx 2.8284$, which has been systematically confirmed in the Bell-type tests on physical entities performed so far \citep{aspectdalibardroger1982,vienna2013,urbana2013}.\label{cirelsonbound}} This can be considered as a convincing argument that the concepts {\it Animal} and {\it Acts} `entangle' whenever they combine to form {\it The Animal Acts}. This entanglement is due to the fact that the combination {\it The Animal Acts} carries meaning in such a way that, depending on which animals and which acts are considered in the human conceptual realm, the majority of micro-states are attributed intrinsically to the combination and not to the individual animals or the individual acts, and hence are not product states but entangled states. As a consequence a classical probabilistic formalism fails, but we have instead proved that a quantum representation in Hilbert space faithfully reproduces all empirical data. The presentation of the technical details would lead us too far from the scopes of the present article. In this regard, we refer the interested reader to \citet{aertssozzo2011,aertssozzo2014,beltrangeriente2018,aertsetal2019,arguellessozzo2020,aertsetal2021,aertsetal2023a}. We instead intend to focus on one specific aspect of the quantum representation, namely, the explicit calculation of von Neumann entropy, as follows. 
\begin{table}
\centering
\caption{We report the data we collected on experiment $AB$ in the five empirical Bell-type tests of entanglement we performed between 2011 and 2021. The table also indicates the von Neumann entropy corresponding to experiment $AB$ (the entropies associated with the other experiments show a similar pattern) and the CHSH factor.}
\label{animalactsentanglements}
%\begin{small}
\begin{tabular}{lllllll}
\hline
Experiments & \multicolumn{4}{l}{Probabilities} & Entropy & CHSH factor \\
\hline
\multicolumn{7}{l}{2011 cognitive test} \\
 & $p(HG)$ & $p(HW)$ & $p(BG)$ & $p(BW)$ & $S$ &  $\Delta_{CHSH}$  \\
$AB$ & $0.049$ & $0.630$ & $0.259$ & $0.062$ & $0.177$ & $2.4197$ \\
\hline
\multicolumn{7}{l}{Google Books test} \\
 & $p(HG)$ & $p(HW)$ & $p(BG)$ & $p(BW)$ & $S$ &  $\Delta_{CHSH}$  \\
$AB$ & $0$ & $0.6526$ & $0.3474$ & $0$  & $0.280$ & $3.41$ \\
\hline
\multicolumn{7}{l}{COCA test} \\
 & $p(HG)$ & $p(HW)$ & $p(BG)$ & $p(BW)$ & $S$ &  $\Delta_{CHSH}$  \\
$AB$ & $0$ & $0.8$ & $0.2$ & $0$ & $0.217$ & $2.8$ \\
\hline
\multicolumn{7}{l}{Google Images test} \\
 & $p(HG)$ & $p(HW)$ & $p(BG)$ & $p(BW)$ & $S$ &  $\Delta_{CHSH}$  \\
$AB$ & $0.0205$ & $0.2042$ & $0.7651$ & $0.0103$ & $0.202$ & $2.4107$ \\
\hline
\multicolumn{7}{l}{2021 cognitive test} \\
 & $p(HG)$ & $p(HW)$ & $p(BG)$ & $p(BW)$ & $S$ &  $\Delta_{CHSH}$  \\
$AB$ & $0.0494$ & $0.1235$ & $0.7778$ & $0.0494$ & $0.114$ & $2.79$ \\
\hline
\end{tabular}
%\end{small}
\end{table}

In an empirical Bell-type test on human participants, a questionnaire is submitted to all participants which contains an introductory text explaining the conceptual entities and precise tasks involved in the test. Let us focus on one experiment for each of the five empirical tests, namely, experiment $AB$ (similar calculations and results hold for the other experiments). Experiment $AB$ can be realised by considering two examples of {\it Animal}, namely, {\it Horse} and {\it Bear}, and two examples of {\it Acts}, namely, {\it Growls} and {\it Whinnies}. Then, the four possible outcomes of $AB$ are obtained by juxtaposing words, so that we get the four options {\it The Horse Growls}, {\it The Horse Whinnies}, {\it The Bear Growls}, {\it The Bear Whinnies}. Next, each participant has to choose which one among these four options the participant considers as a good example of {\it The Animal Acts}.\footnote{In an empirical test on the web, we have instead calculated the relative frequencies of appearance of the four strings of words {\it Horse Growls}, {\it Horse Whinnies}, {\it Bear Growls} and {\it Bear Whinnies}. The way in which tests are performed does not however play any role in what follows.} We denote by $p(HG)$, $p(HW)$, $p(BG)$ and $p(BW)$ the probability that {\it The Horse Growls}, {\it The Horse Whinnies}, {\it The Bear Growls} and {\it The Bear Whinnies}, respectively, are chosen in a given test. These probabilities can be computed from empirical data and are reported in Table \ref{animalactsentanglements} for all empirical studies.

Before coming to the quantum representation, it is important to clarify two aspects that arise from the analysis of empirical data and could be concerning for quantum physicists.

Firstly, the data on {\it The Animal Acts} violate the `no-signalling conditions' \citep{ballentinejarrett1987}. Let us explicitly write one of these conditions (the others are obtained in a similar way). In the case of the coincidence experiments $AB$ and $AB'$ and the outcome $A_1$ of experiment $A$, the probabilities of $AB$ and $AB'$ have to be such that $p(A_1B_1)+p(A_1B_2)=p(A_1B_1')+p(A_2B_2')$. The no-signalling conditions are satisfied in quantum physics, because only coincidence experiments that are represented by tensor product observables are considered. Then, most quantum physicists believe that a violation of these conditions entail the possibility to send `faster than light signals', thus violating Einstein's relativity theory \citep{salartetal2008}. We believe that this conclusion is not correct, as it makes a too quick and superficial transition from, on the one hand, the experimental verification of the existence of the correlations, to a specific nature of these correlations, as if from their nature they could be used to send signals, when this no-signalling condition is not fulfilled. Indeed, even if correlations observed in a Bell-type test are associated with space-like separated events, the mechanism for using them to send signals will not necessarily lead to the type of signal underlying that mechanism that corresponds to what one intuitively imagines, namely, a direct causal signal from Alice to Bob or from Bob to Alice. In \citet{aertsetal2019} we give several examples of macroscopic mechanical situations where (i) Bell's inequalities are violated, (ii) the no-signalling conditions are violated, and (iii) the correlations that are the origin of it belong to space-like separated events. Thus, if causing instantaneous correlations would be sufficient to send signals travelling faster than light, then the macroscopic situations we consider in \citet{aertsetal2019} should also give rise to signals faster than light, which in an obvious way is not the case. It can be readily shown that the presence of correlations of the second kind \citep{aerts1990}, namely correlations that are only potentially present before the coincidence experiment, and are actualized during that experiment, can easily generate space like separated correlated outcomes without a signal taking place faster than light \citep{aertsetal2019,aerts1990}. Hence, the violation of the no-signalling conditions in {\it The Animal Acts}, but also in quantum physics tests, is non-problematical. More, a violation of the no-signalling conditions can be naturally explained if one represents coincidence experiments by means of `entangled observables' \citep{aertsetal2019}.

Secondly, in some Bell-type tests, e.g., the `Google Books' test, the violation of Bell's inequalities also exceeds the known `Cirel'son bound'  \citep{cirelson1980}, that is, the limit of $2\sqrt{2} \approx 2.83$ for the CHSH factor in Equation (\ref{CHSH_factor}), see footnote \ref{cirelsonbound}. Cirel'son bound is usually considered as a theoretical limit to represent the correlations observed in a Bell-type test in Hilbert space. As above, this bound was derived under the assumption that coincidence experiments are represented by tensor product observables, overlooking the fact that, though coincidence experiments are performed in space-like separated regions, this does not imply that experiments should be separated too. Indeed, if coincidence experiments are represented by entangled observables, then the CHSH factor in Equation (\ref{CHSH_factor}) is bound between $-4$ and $+4$, hence a mathematical representation in Hilbert space, using, let us repeat ourselves, both entangled states and entangled observables, becomes possible. Indeed, we have proved in various papers that a Hilbert space representation exists for all data on {\it The Animal Acts}, including those data that also violate Cirel'son bound (see, e.g., \citet{aertsetal2021,aertsetal2023a}). We would like to note that we believe there is more to the Cirel'son constraint that emerges for quantum theory in physical reality, and is readily violated in human cognition, think especially how large this violation is for the data coming from `Google Books' and `COCA'. Our conjecture is that this Cirel'son constraint is connected to an additional symmetry present only in the micro-world and connected to the problematic relationship that the Hilbert space formalism has with separated entities \citep{aerts1982}.

Let us now come to the quantum representation of experiment $AB$ in the combination {\it The Animal Acts}. Since $AB$ has four possible outcomes, {\it The Animal Acts} should be represented, as an overall entity, in the Hilbert space $\mathbb{C}^{4}$ of all ordered 4-tuples of complex numbers. On the other side, a representation of {\it The Animal Acts} as a combination of the entities {\it Animal} and {\it Acts} requires the composed entity to be represented in the tensor product Hilbert space $\mathbb{C}^{2} \otimes \mathbb{C}^{2}$, where an isomorphism $I_{AB}$, which can be taken as the identity operator, maps an ON basis of $\mathbb{C}^{4}$ into the canonical ON basis $\{ (0,1)\otimes (0,1), (0,1)\otimes (1,0), (1,0)\otimes (0,1), (1,0)\otimes (1,0) \}$. Analogy with quantum physics suggests that the composite bipartite entity {\it The Animal Acts} is initially in the pure maximally entangled state represented by the unit vector
\begin{equation}
|\Psi_{12}\rangle=\frac{1}{\sqrt{2}} \Big [ (0,1)\otimes (1,0)- (1,0)\otimes (0,1)  \Big ]
\end{equation} 
in this canonical ON basis. This state corresponds to the singlet spin state and is equivalently represented by the density operator $\rho_{12}=|\Psi_{12}\rangle \langle \Psi_{12}|$. However, whenever the participant starts reading over the introductory text of the questionnaire, this can be considered as a `contextual effect' which changes the initial state into a new generally entangled state represented by the unit vector
\begin{equation}
|\Psi_{AB}\rangle = \sqrt{p(HG)}|HG\rangle + \sqrt{p(HW)} |HW\rangle + \sqrt{p(BG)} |BG\rangle + \sqrt{p(BW)} |BW\rangle
\end{equation}
where $\{|HG\rangle, |HW\rangle, |BG\rangle, |BW\rangle\}$ is an ON basis of eigenvectors of the product self-adjoint operator which represents $AB$ in $\mathbb{C}^{2} \otimes \mathbb{C}^{2}$. To calculate the von Neumann entropy, we write $\rho_{AB}=|\Psi_{AB}\rangle \langle \Psi_{AB}|$ and get $S(\rho_{AB})=0$, since $\rho_{AB}$ represents a pure state.

Let us now calculate the von Neumann entropies associated with the states of the component concepts {\it Animal} and {\it Acts} by taking the partial traces of $\rho_{AB}$ with respect to {\it Acts} and {\it Animal}, respectively. One finds \citep{aertsbeltran2022a} that
\begin{equation}
\rho_{{\rm Animal}} = Tr_{\rm Acts} \rho_{AB}=
\begin{pmatrix}
p(HG) + p(HW) & \sqrt{p(HG)p(BG)} + \sqrt{p(HW)p(BW)} \\
\sqrt{(p(BG)p(HG)}+ \sqrt{p(BW)p(HW)} & p(BG) + p(BW) 
\end{pmatrix}
\end{equation}
and 
\begin{equation}
\rho_{{\rm Acts}} = Tr_{\rm Animal} \rho_{AB}=
\begin{pmatrix}
p(HG) + p(BG) & \sqrt{p(HG)p(HW)} +  \sqrt{p(BG)p(BW)} \\
\sqrt{p(HW)p(HG)} + \sqrt{p(BW)p(BG)} & p(HW) + p(BW) 
\end{pmatrix}
\end{equation}
Then, we diagonalise the self-adjoint operators $\rho_{{\rm Animal}}$ and $\rho_{{\rm Acts}}$ and use Equations (\ref{vonnneumann}) and (\ref{shannon}) to calculate the corresponding von Neumann entropies. The results are reported in Table \ref{animalactsentanglements}. As we can see, the von Neumann entropy of both {\it Animal} and {\it Acts} is greater than zero in all studies, which is consistent with the fact that the concepts {\it Animal} and {\it Acts} are in non-pure, or density, states. Hence, we have obtained an important result: if one considers a very simple text which is the combination of just two words, namely, the words corresponding to the concepts {\it Animal} and {\it Acts}, 
the entropy of the overall text corresponding to {\it The Animal Acts} is less than the entropy of the component words corresponding to {\it Animal} and {\it Acts}. Equivalently, the `process of conceptual combination, or composition, reduces the entropy of the component entities'. 

We believe that the result above has a general validity in language: if we consider any text, together with its meaning content, then it is reasonable to assume that it is in a pure entangled state, because of the way `meaning' is carried by the combination, giving preference to entangled micro-states over product micro-states, we refer to our analysis about the cat and dog example in Section \ref{energy}. As such, this entangled state will have a von Neumann entropy that is equal to zero. However, this entropy will be less than the von Neumann entropies of the component words. This feature that the von Neumann entropy of the composed entity is systematically smaller than the von Neumann entropies of the component entities, i.e. `decreasing entropy as a consequence of composition' is completely new, as it is not present in classical thermodynamics where instead `entropy tends to increase as a result of composition'. Indeed, a classical thermodynamic framework would require the overall text to be, at equilibrium, in the state that maximises the total entropy. In other words, while classical (thermodynamic or information) entropy is typically associated with the `disorder' of the composed entity, von Neumann entropy is typically associated with the `order', meant in the sense of more fine-tuned meaning, of the composite entity.  

Another important conclusion that can be drawn from the analysis above is that, at a conceptual level, the words of a text in an entangled state in some way `lose their identity', exactly as it occurs in physics whenever quantum entities entangle. When words or concepts entangle, we understand what happens: words and concepts have the fluidity and contextuality to serve well and with meaning content (changed but not disappeared) whenever they combine.

We will see in the next section that this non-classical feature of entropy, together with the non-classical features of energy (see Section \ref{energy}), suggest that a thermodynamic theory which relies on the notions of energy and entropy can be developed for language, and provide arguments to suggest that it can be extended to human cognition and some broad areas of human culture. But, this has to be a non-classical quantum-type thermodynamics.

\section{Towards a non-classical thermodynamics for human cognition and culture \label{thermodynamics}}
In Sections \ref{energy} and \ref{entropy}, we have studied concrete situations of human language where we have highlighted the relationship between words and texts, in particular, their relationship in terms of meaning and content. The results obtained in those sections in specific cases allow one to put forward `important points' with respect to the fundamental characteristics of a thermodynamic theory of language. Indeed, those results significantly show that (i) the notions of energy and entropy can be consistently introduced and also precisely quantified in human cognition, hence (ii) a thermodynamics language can be worked out starting from the notions of energy and entropy, exactly as classical thermodynamics, as a physical theory, has been worked out on these two notions. However, the analysis presented in Sections \ref{energy} and \ref{entropy} also reveals that any thermodynamic theory of language has to be definitely a `non-classical theory', because energy and entropy exhibit `systematic and genuine quantum-type aspects', as we have illustrated. 

More precisely, we have proved that the energies of words in a large text statistically behave in a way that does not satisfy Maxwell-Boltzmann but, does satisfy Bose-Einstein, statistics. This means that, due to the fact that the concepts associated with the words that appear in any text carry meaning in such a way that preference is given to entangled micro-states to product micro-states,  the quantum  aspects of energy quantization, coherence, indistinguishability and Bose-Einstein condensation are at place to describe well the dynamics of this meaning. Furthermore, we have proved that the entropy of words in a text decreases as a result of their composition, or combination. This means that, again due to the fact that the concepts associated with the words that appear in any text carry meaning in this way, the quantum aspects of entanglement, non-separability and contextuality are at place to describe well the entropy of the overall text, including it to be lower than the entropy of the component words, which is exactly the opposite than the behaviour of (thermodynamic or informational) entropy of classical composite entities.

In other words, the lesson one can learn from the analysis in the present article is the following: classical thermodynamics rests on the intuition that a gas consists at the microscopic level of entities that are random, and maximisation of entropy expresses this intuition. In the case of cognition, we have seen that there is no situation where micro-level words are random. In fact, our explanation for the Bose-Einstein behaviour, i.e. `bundling of words in a text', is that this phenomenon is due to the meaning carried by the text. Then, it should not be a surprise that there is a decrease in entropy, as it follows from the von Neumann entropy calculations we have made in Section \ref{entropy}, and thus as a consequence of entanglement. The intense concentration required by a human mind to write a well-formulated text is probably due to causing this decrease in entropy, and the creation of entanglement is then probably a `tool' wielded with that intention.  

We conclude the article with two important remarks which can widely extend the scale of the results obtained here. Indeed, we believe that the genuine quantum aspects identified in our investigation, in particular, quantization of energy and entanglement-induced reduction of entropy, are not peculiar of language, but they generally hold in human cognition and also in human culture, for example, with cultural artefacts. Let us illustrate what we mean by simple examples involving cognition and cultural artefacts. 

Let us firstly consider the phenomenon of `categorical perception’, which is systematically present in cognition, but also, more generally, in many areas of human activity \citep{goldstonehendrickson2010}. The phenomenon consists in the experimental finding that our perception, e.g., spatial perception, is `warped' so that differences between stimuli we classify in different categories are magnified, while differences between stimuli we classify in the same category are reduced. Our perception of colours is a good example of this phenomenon. We perceive a discrete set of colours, more specifically, the colours of the rainbow, though in the physical reality there is a continuum of different frequencies over the frequency range of visual light offered to us as stimuli. Thus, the warping effect of the categorical perception of colours consists of the fact that two stimuli that both fall within the category of `green' are perceived as more similar than two stimuli of which one falls within the category of `green' and the second within the category of `blue' even if, from a physics perspective, both pairs of stimuli have the same difference in frequencies. The simultaneous presence of these two effects, a contraction within an existing category, and a dilation between different categories, causes a clustering together into clumps of colours, which in the end gives rise to the seven colours of the rainbow. We believe that this discretization that is present in categorical perception can be naturally explained as a phenomenon of quantisation of energy, exactly as it happens in the quantization of light  (and exactly as it happens in language). We have not quantitatively investigated the phenomenon yet, but our results in language indicate that one should expect the presence of quantum-type indistinguishability and Bose-Einstein statistics. A formal analysis of the phenomenon has been presented in \citet{aertsaertsarguelles2022}.

Secondly, every collaboration between humans exploits the effect we have identified in Section \ref{entropy}, namely, the creation of entanglement correlations in any collaboration of different individuals which lowers the entropy of the collaboration as compared to the entropies carried by every individual who participates in the collaboration. Let us consider, e.g., the act of `hunting'. Suppose a given number of hunters have to collaborate to make a successful hunting in a forestry. The behaviour of the prey is unpredictable, e.g., the animal may suddenly run away, or move towards one of the the hunters, who then has to hide behind a tree, etc. Hence, there is an element of uncertainty accompanying the overall process. In addition, each individual, taken separately from the others, has in principle a lot of freedom too, which is meant to be used in function of making the collaborative hunt as little uncertain as possible, hence to bring its state as a close as possible to be a pure state, or its entropy as little as possible. On the individual level of each hunter however even during an almost perfect collaborative hunt with very low entropy, a lot of uncertainty is needed to be left open as a potential to be used by the individual in function of the maximisation of the certainty of the collaborative hunt. This literally means that human collaboration strives at minimizing entropy of the collaboration exploiting the tool of entanglement within the collaboration and the presence of a quantum-type thermodynamics in human culture to allow for this mechanism. The many macroscopic material entities we proposed in our Brussels group that violate the Bell inequalities illustrate very well this mechanism, of how `human cooperation' introduces entanglement between the various individual participants resulting in an entropy reduction for the whole, the examples are proposed as experiments to be performed in a lab situation, but always involve at least two cooperating experimenters, Bob and Alice \citep{aertsetal2019}. The question arises, e.g., in connection with an example such as that of `the hunt', where are the quanta and energy levels? Our answer is that we can rightly put forward human language as the most advanced substrate of human culture and human cooperation, where the quanta are clearly demarcated, namely, the concepts, and that in other less delineated portions of human cooperation this quantization is  still less evolved, hence less structured, and also less noticeable but still present. With respect to `the hunt', we can think of quanta that are standardized pieces of movement, tactical interventions, strategies, as they exist very explicitly in current human sports, for example, which can be considered as more evolved versions of the more ancient forms of collaboration, such as `the hunt'. The thermodynamic theory for it can be identified exactly like we approached it for human language, hence the more frequent of such standardized piece of movement the lower its energy level. It seems a very unexplored area of experimental psychology needed to build out our thermodynamic theory for these quantized areas of human culture, and therefore that we note here additionally that much of it has already been done by researchers studying the phenomenon of categorical perception. We give a different example than this of the colors that we cited above, and which will be more surprising to those who are not familiar with the phenomenon of categorical perception. It was shown that `emotions' are quantified in six different facial expressions. And a hint of where our thermodynamics for human culture can first be systematically developed is given, as we explained in the previous section, by the many experimental investigations that identify the validity of Zipf's law in human culture.

To conclude, these examples, together with Zip's law being present in many areas of human activity, suggest that the non-classical behaviour that we have identified in this article is not limited to language or cognition, but also concerns human culture. And, more, we believe that the reduction of von Neumann entropy as a consequence of composition, or combination, is the crucial ingredient to also build a thermodynamic theory of quantum physics, which is being developed but does not meet full agreement of the scientific community on how it should be  understood.

\end{document}